# Truth, Beauty and Supergravity

S. Deser
Walter Burke Institute for Theoretical Physics,
California Institute of Technology, Pasadena, CA 91125;
Physics Department, Brandeis University, Waltham, MA 02454
deser@brandeis.edu

Abstract
I use Supergravity as a test case to study the role and uses of elegance/simplicity in formulating and evaluating physical models, whose sole criterion is of course "truth"— an observationally verified description of Nature within a certain range of scales.

Introduction
I begin with a warning: theoretical physics is an edifice built over the centuries by some of mankind's greatest minds, using ever more complicated and sophisticated concepts and mathematics to cover phenomena on scales billions of times removed—in directions bot bigger and smaller—from our human dimensions, where our simple intuition or primitive language cannot pretend to have any validity. So any popular discussion is necessarily impressionistic, being couched in terms of classical analogies that do not really apply. This warning label should be attached to all such accounts, the present one included. However, I have tried to focus here on an aspect that involves more human attributes of our subject.

All theoretical physicists sooner or later grapple with is the role of beauty, often also called elegance, in confirming the correctness of our natural laws. In part, this is a problem of language: any "good" theory acquires beauty as its correctness is confirmed—we find hidden aspects to marvel at. Conversely, those models that do not seem to be used by Nature despite their apparent formal attractions eventually lose their luster. Yet there is some deep sense in which the two —truth and beauty —are linked. Among our great scientists, the range goes from Boltzmann who said "Eleganz ist fuer Schneider", elegance is for tailors, to Einstein for whom beauty would force the Lord to accept a theory, despite apparent experimental contradictions, as was eminently the case for Special —and to a lesser extent—General—Relativity, and even more so in the recent history of our "standard model" of the basic microscopic laws of Nature. Newton's famous remark about picking up pretty pebbles at the seashore, instead of facing the vast ocean of truth lying just beyond, seems to exhibit a more ambivalent attitude.

I chose Supergravity (SUGRA), to illustrate this topic because it is of one of our most recent and conceptually most novel entries: it is just over four decades old, and already has a literature of about 15,000 papers! This model, along with its wider, and also recent, ancestor, Supersymmetry (SUSY), thus provides a perfect, fresh, case study. SUGRA also covers a broad canvas, including General Relativity, Quantum Field Theory and their unification, which is currently our subject's holy grail. I shall of course avoid scaring you, with long—indeed, with any —formulas; yet I emphasize that, as Einstein said, things should be stated as simply as possible, but not more simply!

## The components
To set the background, one of the equations most perfectly beautiful and most perfectly in accord with Nature is the Dirac equation governing the behavior of electrons—as well as all the other leptons and quarks, hence also our protons and neutrons. Indeed, it is perhaps one of our three most beautiful equations, along with Maxwell's and Einstein's! It came full-blown from the head of one of the true greats of the last Century, and instantly divided all particles into two antipodal types, the Bosons, e.g., mesons and photons that like to congregate (think of intense laser beams) and the Fermions that hate to do so to the maximal extent (Pauli's exclusion principle). Like Greta Garbo, they want to be alone, while Bosons are gregarious, more like…Mae West.

Once invented, any interesting equation in physics is just asking to be generalized, and there are always people willing to oblige. In the Dirac case, his equation—which only makes sense at the quantum, rather than classical, level—describes particles with intrinsic spin, like little tops, but the spin is necessarily fixed to be one half unit of the basic value, namely the famous Planck constant that started all Quantum theory off back in 1900. The next possible allowed values for a fermion would be 3/2, 5/2... units. The fact that no such elementary particles had ever been seen was no obstacle, and in due course the counterpart of the Dirac equation for spin 3/2 was produced, both endowed with mass and without—it is the latter we shall use. Indeed mass and spin are the two intrinsic basic parameters that label any particle or field (the two words are interchangeable in the quantum world). There matters rested until the "super" revolution began to take hold in the early nineteen seventies.

At this point, I must remind you a bit about General Relativity (GR). Its absolutely novel point was to make Geometry, so familiar for millenia as the passive theater in which matter interacts, into a dynamical ever-changing entity of its own, subject to laws of motion—here the Einstein equations— rather than fixed once for all by fiat—indeed geometry was (almost) the last "a priori" to fall; why our space-time has dimension 4 is still the exception! Those Einstein equations specify how geometry reacts to—and determines the course of— matter, that is all other fields. Further, it is a universal notion in that all matter must interact uniformly with gravity: none is exempt and indeed Geometry reacts with itself in a complicated (geometrical ) way. All these crazy-seeming ideas have observational consequences that include Newton's old Universal law of gravitation but in a far more coherent and general way, with predicted corrections that have always been verified and never contradicted. The latest, truly spectacular, triumph involves the (now several) observations of gravitational waves—incredibly tiny spacetime oscillations that were predicted a century ago when GR was invented, but only observed in the past year by the unbelievably refined laser beam detectors of LIGO; even better, these waves could be traced back to another crazy prediction of GR, namely Black Holes, whose collisions emit them as debris. So we can certainly believe Einstein's theory, though as with any theory, only at the scales where it has proved reliable—here a pretty hefty stretch from the (pretty small) to the very big.

Let me also remind you of Einstein's (breathtaking) dream, that of unification of geometry and matter into a unitary whole. This dream became his late years' obstinate, but fruitless, quest, although it did lead to many unexpected new concepts: in particular that our Universe may exist in more than four dimensions; this in turn became an essential aspect of String theory. I should end this resume of GR by noting that gravitational waves consist, at quantum level, of Bosonic particles that we call gravitons, just like the familiar electromagnetic spectrum is made of photons, also Bosonic, that is integer spin, with respective spins (2,1). Furthermore, GR has the dual quality of also being expressible as a theory of "normal" matter, in which its geometrical aspects are exchanged for a well-understood dynamical matter-like description. Indeed, GR is as beautiful In this dual way as it is geometrically.

So here is pure geometry on the one hand and brute matter on the other, in particular those strange, but essential Fermions of which we are made. Surely unification could never wed these antipodal concepts, or could it? This is the realm of our newest playground, SUSY, discovered in Moscow in 1969 and independently in New York a few years later, and also traceable to early string theory. I must now give you a few words about this—yes, extremely beautiful by unanimous physicists' consent—concept.

Let's take a step back: Historically, the greatest progress in physics was the notion of invariance under some set of transformations—think of the most elementary: rotations and translations in our ordinary Euclidean three-dimensional space—the world still looks the same even if you move uniformly in some fixed direction (per Galileo) or turn your chair to another angle. This notion can be generalized to more abstract spaces, but with the same underlying idea. The spaces may be labelled by some properties of a set of particles, all of which behave similarly under various interchanges between them in certain contexts. So SUSY would put Bose & Fermi particles on an equal footing in certain "rotations", without taking away their distinctive crowd behaviors of being gregarious or the opposite. Mathematically, it was a small step, which however has now generated an absolutely enormous physical and mathematical literature. Indeed, the LHC accelerator at CERN was designed not only to seek (& found) the "Higgs" (spin 0) boson, but also to find traces of SUSY, that is, companion particles of the known particles, but with opposite "polarity". That's a lot of hard lore to digest, but just think of the rotation invariance analogy, in which the angle of rotation represents mixing of the x- and y-axes here represents mixing Bosons & Fermions, of adjoining spin like 1/2 and 1—Dirac-like particles and photon-like ones. That's it—"Reader's Digest" SUSY!

Elegance of the other pillar of fundamental physics, the "standard model" describing all known matter in a unified way, is a mixed bag, while being an absolutely correct and universal—as measured to date—"true" description of matter's behavior. We have come to love it for that, but not for its some twenty free parameters nor for its seemingly haphazard cascade of invariances—we sympathize with the eminent elder statesman Isidore Rabi, when he exclaimed about an especially odd new particle, "who ordered that?" Yet there has never been a truer and as encompassing an edifice as the standard model.

Unification and its discontents

So here we (almost) are, trying to make the most elegant of all theories, unifying Einstein's and (generalized) Dirac's equations, a combination of adjoining spins (2, 3/2) that cries out to be joined. The payoff is nothing less that, as mentioned, the eternal dream of unifying geometry— that is space, with not just any matter, but Fermionic matter, at that! Indeed, in a technical sense, the Dirac part would be the (spinorial) square root of gravity. To spare you the suspense, this attempt was successful—made independently and simultaneously—just 41 years ago, by two separate groups [1]. Actually, SUGRA is even more beautiful that mere SUSY, because it enjoys a much deeper "local rotation" invariance. Even more serendipitously, the combined equations governing it are the simplest possible, with the least tricky, "minimal", possible interaction.

But as the TV advertisers say, "wait, there's more—lots more", In fact maximally more. It turns out that once we have linked their adjoining spins, the game can be continued to include spins 1, 1/2 and 0, i.e., all possible spins from 2 down can play. Indeed, we even understand why (elementary) spins bigger that 2, such as 5/2 or 3, are forbidden, as they are in Nature as well—there's simply no consistent room for them. Still further, we can extend our search to higher dimensions than 4, all the way to 11 in fact, beyond which no SUGRAS with maximal spin 2 can be constructed at all (recall that, coincidentally(?), superstrings live in 10 dimensions). And that's still not all! I have yet to mention perhaps the strongest motivation for SUGRA(S), their quantum behavior. The most burning problem in our physical theories is believed to be the very bad behavior of GR when we attempt to quantize it and study the consequences: Unlike all other fields that make up our Universe, gravitons give rise to uncontrollable ultraviolet (UV) infinities as the answer to any physical questions, thereby negating any predictive power, at least in the step-by-step, "perturbative", regime that is the only one we know how to use. So any improvement of this UV catastrophe could be a strong hint as to how to cure it. In fact, SUSY's original attraction was due in great part to the cancellation of infinities in its various models between the Bosonic & Fermionic components.

To what extent these miraculous cancellations extend to SUGRA is the obvious big question. The answer is rather mixed: original SUGRA, as described above—perhaps the simplest and most beautiful model—in fact stays UV finite at the first two orders of perturbative calculation, for deep invariance reasons to boot, unlike any other GR+matter system! All this for the original SUGRA, as I said. However, the SUGRA industry was busy generalizing this model to include all lower spins down to 0, the most complete one being the so-called N=8 model, the original being N=1, and GR itself "N=0". Indeed, there has grown an enormous industry—unsurprisingly, given the stakes—to calculate higher than second approximation. The complication grow so rapidly with higher orders that it is not yet quite clear whether things stay finite even to all single digit orders, already a truly enormous enterprise. Of course the world we see is certainly neither SUSY nor SUGRA, but at best "broken", for example because these new companion particles, if they exist at all, have non-0 mass, unlike the ones in SUGRA, so even perfect finiteness would still be only a hint, one whose import we do not (yet?) understand.

Conclusion

Now comes the time for the punchline—wise general remarks regarding truth and beauty in physics, as exemplified by SUGRA. Let's summarize what we have described so far. There are certain ideas, equations and theories in physics that are almost universally recognized by its practitioners as beautiful and elegant. This may occur quite independently of their empirical or observational verifications; indeed it often occurs despite the apparent clash between their predictions and experiment. We emphasized that this was the case for some of the most sublime—and later vindicated—such as GR and the Dirac equation. Of course the eye of the beholder is conditioned by education, experience and the collectively accepted state of the art, all rather subjective criteria: Newton could not have directly understood the wonders of Dirac's or Maxwell's or Schrodinger's equations (although he would have caught on with some help, and then surely agreed). That it requires a trained practitioner to appreciate the lightning stroke of a new creation holds true for the arts as well. It is perhaps more surprising to the usual popular image of the scientist that elegance and beauty play such leading roles, and it must also be admitted, as I mentioned, that a concept that provides widespread empirical unification will thereby acquire esthetic value, simply from the many unexpected facets its usage uncovers.

Our chosen example, SUGRA, certainly qualifies on the elegance and beauty scales, if only because of its parent theories, Einstein and Dirac. Right from its birth, it felt like a new art form. On the truth front, however, it's been another story altogether: no elementary spin 3/2 fermion has ever been found, even in some implicit way, nor indeed any of the companion particles predicted more broadly by SUSY. They could be lurking just outside the range of LHC or current cosmological observations. But at present, it must be acknowledged that there is no evidence at all that Nature agrees with the esthetic appeal. And I must emphasize that in the end, if the next scale our instruments can probe still fails to find them, they may exemplify Boltzmann's dictum that only tailors would find SUGRA compelling. Yet, at the very least, important theoretical advances have been made in our understanding of pure GR, just by knowing that it can be unified with spin 3/2 matter, whether or not it is so unified!

Truth, by contrast with beauty, would seem to be a far simpler, more direct, aspect of physics: after all, when a theory is verified to many decimal places ( as many as 12 in some cases!) in widely different areas, it hardly seems worth even questioning its "truth". Yet, here too, things are far less simple than they would seem. One example is the theory known a QED, Quantum Electrodynamics, the basis for all atomic phenomena, that occupied much of 19th and over half of 20th Century experimental and theoretical research. It is unsurpassed, of all human endeavors, in the accuracy and correctness of all its predictions (those 12 decimal places), is certainly beautiful and simple to state (being the quantum expression of the Dirac plus Maxwell equations), but is equally certain to be wrong at a more fundamental level:  When pushed too far, it is revealed to be full of internal contradictions and loss of predictive power. Yet there is no doubt whatever that its incredibly accurate predictions in its domain of validity are entirely valid and reliable! On the other hand, a recent extension of QED, called QCD for Quantum Chromodynamics, reigns unchallenged in explaining the subnuclear domain governing quarks in the standard model. It is fully as beautiful as QED, although initially regarded as a bit of an ugly duckling, even by its discoverers. Its most basic "prediction", confinement, that we believe makes quarks condense permanently into our protons and neutrons, has never yet been entirely proved, nor has it ever been seriously doubted!

I don't mean to exhibit these (only apparent) pathologies in our physics thinking in a pejorative way: more to give a flavor of what the elaborate work of many physicists over lifetimes has been distilled to. Physics is most likely a never-ending quest, not just in the poetic sense, but literally according to the beautiful concept I end with: Kenneth Wilson's (and others') ideas of the unfolding of novel conceptual aspects of the universe as one widens the scale of enquiry, all in a very concrete well-founded sense [2]. That is perhaps the most beautiful idea of all!

Acknowledgements.
This work was supported by grants NSF PHY-1266107 and DOE#desc0011632. An earlier version appeared on the PRC Mr Science program; I thank A.Zee for inviting it.